\documentclass{proceedings_style}

\pdfoutput=1

\usepackage{graphicx}
\usepackage{amsmath}
\usepackage{amssymb}
\usepackage{float}
\usepackage{appendix}
\usepackage{listings}
\usepackage{subcaption}
\usepackage{cite}

\newcommand{\stringa}{\ttfamily\lstinline}
\def\cod#1{{\stringa!#1!}}
\def\ie{{\it i.e.}}

\title{Multi-jet production in the high energy limit at LHC}

\ShortTitle{Multi-jet production in the high energy limit at LHC}

\author{F. Caporale\\
        Instituto de F{\' \i}sica Te{\' o}rica UAM/CSIC, Nicol{\'a}s Cabrera 15\\
        \& Universidad Aut{\' o}noma de Madrid, E-28049 Madrid, Spain.
        E-mail: \email{francesco.caporale@uam.es}}
        
\author{F.~G. Celiberto\\
		 Instituto de F{\' \i}sica Te{\' o}rica UAM/CSIC, Nicol{\'a}s Cabrera 15\\
        \& Universidad Aut{\' o}noma de Madrid, E-28049 Madrid, Spain.
        E-mail: \email{francescogiovanni.celiberto@fis.unical.it}}
        
\author{G. Chachamis\\
        Instituto de F{\' \i}sica Te{\' o}rica UAM/CSIC, Nicol{\'a}s Cabrera 15\\
        \& Universidad Aut{\' o}noma de Madrid, E-28049 Madrid, Spain.
        E-mail: \email{chachamis@gmail.com}}

\author{\speaker{D. Gordo G{\' o}mez} \thanks{La Caixa-Severo Ochoa Scholar.}\\
 Instituto de F{\' \i}sica Te{\' o}rica UAM/CSIC, Nicol{\'a}s Cabrera 15\\
        \& Universidad Aut{\' o}noma de Madrid, E-28049 Madrid, Spain.
        E-mail: \email{david.gordo@csic.es}}

\author{A. Sabio Vera\\
 Instituto de F{\' \i}sica Te{\' o}rica UAM/CSIC, Nicol{\'a}s Cabrera 15\\
        \& Universidad Aut{\' o}noma de Madrid, E-28049 Madrid, Spain.
        E-mail: \email{a.sabio.vera@gmail.com}}

\abstract{
We briefly review observables that were recently proposed in inclusive 3-jet production at hadronic colliders. They are based on azimuthal-angle correlations between the final-state jets and  can be seen as a generalisation of the usual Mueller-Navelet setup. We discuss the stability of the observables once higher order corrections --beyond the leading-logarithmic accuracy-- are taken into account. Finally, we give a status report on a related project in which we examine 
azimuthal angle correlations for Mueller-Navelet jets after imposing a rapidity veto which forbids subsequent 
minijet emissions to be very close in rapidity.}

\FullConference{  Low x Workshop \\
                 June 13-18 2017 \\
                 Bari, Italy}

\begin{document}

\section{Introduction}

Semi-hard processes in Quantum Chromodynamics (QCD) with a strong scale hierarchy allow for 
 the appearance of large logarithms  that could spoil the convergence of the perturbative expansion.
These logarithms are large logarithms in  $s$, the center-of-mass energy squared, and can be resummed 
within the Balitsky-Fadin-Kuraev-Lipatov (BFKL) framework  at leading logarithmic (LLA)~\cite{BFKLLO}
 and next-to-leading logarithmic (NLLA) approximation~\cite{BFKLNLO}.

  One expects the onset of BFKL related effects
 to take place already before we reach asymptotically high energies. However, it is still an open issue 
 whether a clear and undisputed signal of BFKL dynamics may be disentangled at present  colliding energies at the LHC. 
 A promising path in the search for BFKL related effects at hadron colliders is the study of Mueller-Navelet jets (dijets)~\cite{Mueller:1986ey}\footnote{Another idea, suggested in~\cite{Ivanov:2012iv} and investigated in~\cite{Celiberto:2016hae,Celiberto:2017ptm}, is to study the production of two charged light hadrons, $\pi^{\pm}$, $K^{\pm}$, $p$, $\bar p$, well separated in rapidity.}, configurations with two final-state jets ($A$ and $B$) that have similar size transverse momenta $k_{A,B}$ and  large separation in rapidity, $Y=\ln ( x_1 x_2 s/(k_A k_B))$.

A large number of studies in the literature 
\cite{BFKLstudies}~
is focused on the azimuthal angle difference $\theta$ between the two jets. Assuming no minijet radiation, 
one would expect the two jets to have a back-to-back kinematical configuration or, in other words, the two jets to be fully
correlated on the transverse plane. Any azimuthal decorrelation will have to be associated to minijet activity between
the two jets. Within the BFKL framework, the minijet activity can be accounted for by the gluon Green function which allows
for quantitative theoretical predictions regarding the azimuthal decorrelation between the two jets. More interestingly,  it was shown~\cite{Vera:2006un,Vera:2007kn} that ratios of azimuthal angle correlations
\begin{eqnarray}
\label{eq:ratiosMN}
{\cal R}^m_n = \langle \cos{(m \, \theta)} \rangle / \langle \cos{(n \, \theta)} \rangle\,\,,
\end{eqnarray}
(where $m,n$ are positive integers) are much more favourable quantities when searching for the onset of BFKL effects at present phenomenological energies. The comparison of different BFKL NLLA calculations for the ratios ${\cal R}^m_n$
\cite{BFKLNLLAstudies}~
against LHC experimental data at $\sqrt{s}=7\text{ TeV}$ indicates that the analyzed kinematical domain lies in between the regions described by the DGLAP and BFKL approaches~\cite{Khachatryan:2016udy}.

Along these lines, new observables were recently proposed for processes with three~ \cite{3jets}~
or four tagged jets~\cite{4jets}~
 in the final state, with the outermost jets having a large rapidity separation and any other tagged jet restricted in more central regions of the detector. 
Here we will only discuss inclusive three-jet production, assuming that the tagged jets are connected in the $t$-channel via two gluon Green functions. The main idea, presented in Refs.~\cite{3jets},
was to produce theoretical estimates for the ratios 
\begin{eqnarray}
R^{M N}_{P Q} =\frac{ \langle \cos{(M \, \theta_1)} \cos{(N \, \theta_2)} \rangle}{\langle \cos{(P \, \theta_1)} \cos{(Q \, \theta_2)} \rangle} \, = \, \frac{C_{MN}}{C_{PQ}} \,, 
\label{Rmnpq}
\end{eqnarray}
where $\theta_1$ is the azimuthal angle difference between the first and the second (central) jet, while $\theta_2$ is the azimuthal angle difference between the second and the third jet. It is important to note that, contrary to the dijets case, the observables involving sines of azimuthal angle differences do not vanish for events with larger multiplicity. Nevertheless, here we study relations with cosines for simplicity. The next-to-leading logarithm corrections to the gluon Green's function are large and, in principle, they could have a strong impact on the observables $R^{M N}_{P Q}$. Therefore, it is important to assess the stability of the observables after considering the gluon Green functions at NLLA. For a detailed discussion we
refer the reader to~\cite{Caporale:2016zkc}, here we will only show theoretical predictions for the ratio
$R^{12}_{33}$ calculated at NLLA.

We will assume for the outermost jet transverse momenta that fulfil the following limits:
$k_A^{\rm min} = 35$ GeV, $k_B^{\rm min} = 50$ GeV,  $k_A^{\rm max} = k_B^{\rm max}  = 60$ GeV.
The transverse momentum of the central jet, $k_J$, can take values in three wide bins, that is, ~$20\, \mathrm{GeV} < k_J < 35\, \mathrm{GeV}$ (bin-1),
$35 \,\mathrm{GeV} < k_J < 60\, \mathrm{GeV}$ (bin-2) and
$60\, \mathrm{GeV} < k_J < 120\, \mathrm{GeV}$ (bin-3). Each bin corresponds to a central jet with transverse momentum smaller, of the same order, or bigger than the outermost ones respectively.
To quantify the difference between LLA and NLLA, we define 
\begin{eqnarray}
\delta x(\%) = \left(
 \text{res}^{\rm(LLA)} - \frac{\text{res}^{\rm (BLM-1)}+\text{res}^{\rm (BLM-2)}}{2}
 \right) \frac{1}{ \text{res}^{\rm(LLA)}}\,\,,
 \label{corrections}
\end{eqnarray}
where $\text{res}^{\rm(LLA)}$ is the LLA result, while $\text{res}^{\rm(BLM-1)}$ and $\text{res}^{\rm(BLM-2)}$ are NLLA results in the BLM scheme~\cite{BLM} differing in the value of the renormalisation scale in order to give us a measure of the theoretical uncertainty.

\section{Results}
\begin{figure}[H]
\hspace{-.9cm}
\begin{subfigure}{.1\textwidth}
\centering
   \includegraphics[scale=.9]{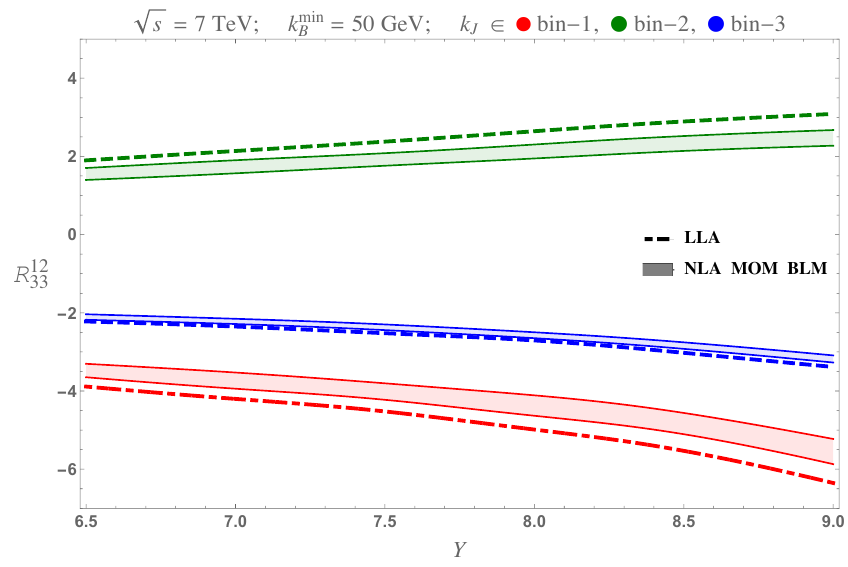}
\end{subfigure}
\begin{subfigure}{1.42\textwidth}
\centering
   \includegraphics[scale=.9]{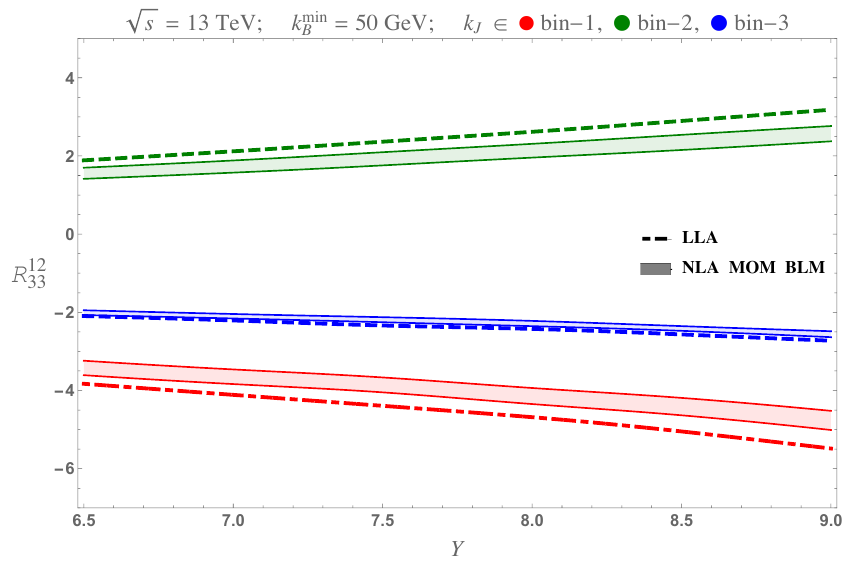}
\end{subfigure}
\caption{\small $Y$-dependence of the LLA (dashed lines) and NLLA (bands) results for $R^{12}_{33}$ with $y_J$ integrated over a central rapidity bin at $\sqrt s = 7$ TeV (left) and $\sqrt s = 13$ TeV (right).} 
\label{fig:first}
\end{figure}

Firstly, we present results for $R^{12}_{33}$ as a function of the rapidity difference of the two outermost jets after integrating over a central jet rapidity bin, \ie, after allowing for the central jet rapidity to take values in the range  $-0.5 < y_J < 0.5$, see Fig.~\ref{fig:first}.
We use dashed lines to represent LLA results and a band for the NLLA results. The band is bounded by the $\text{res}^{\rm(BLM-1)}$ and $\text{res}^{\rm(BLM-2)}$ results in thin continuous lines. 
The red curve and the red band represent the LLA and NLLA results, respectively, for the central jet belonging to bin-1, the green curve and the green band for the central jet belonging to bin-2 and the blue curve and the blue band for the central jet belonging to bin-3. We see that the NLLA corrections are mild. Moreover, we note that the overall picture almost does not change when we go from $\sqrt{s}=7\text{ TeV}$ to $\sqrt{s}=13\text{ TeV}$, except for some minor changes in the large rapidity region, due to the presence of the phase space boundary at $\sqrt{s}=7\text{ TeV}$.

\begin{figure}[H]
\hspace{-.9cm}
\begin{subfigure}{.1\textwidth}
\centering
   \includegraphics[scale=0.9]{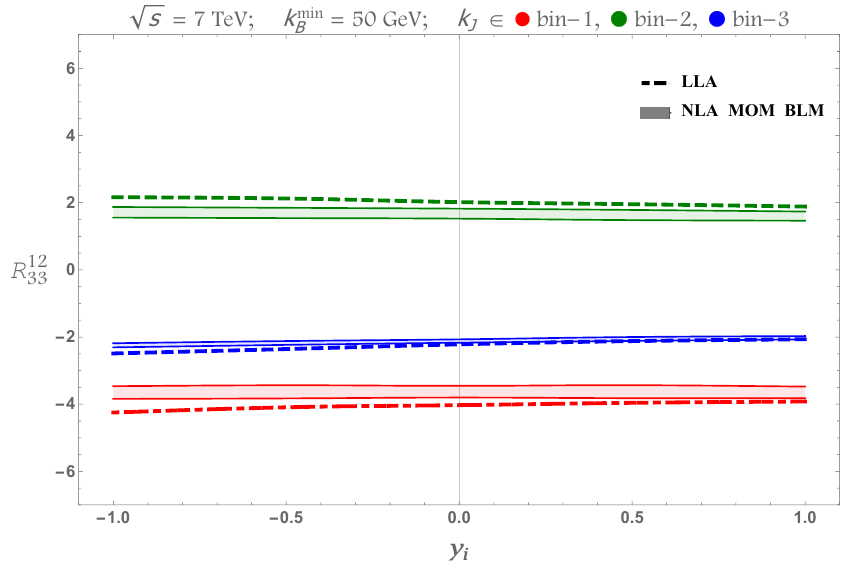}
\end{subfigure}
\begin{subfigure}{1.42\textwidth}
\centering
   \includegraphics[scale=0.9]{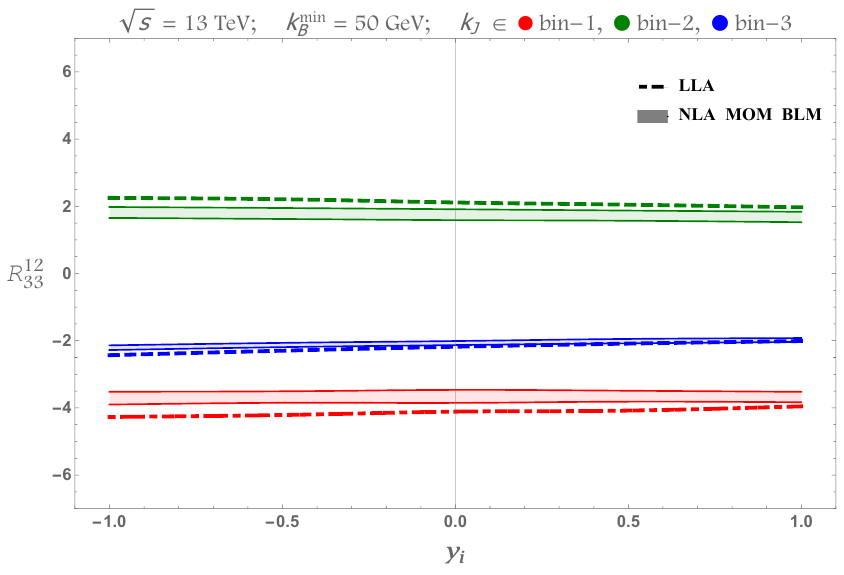}
\end{subfigure}
\caption{\small $y_i$-dependence of the LLA (dashed lines) and NLLA (bands) results for $R^{12}_{33}$  at $\sqrt s = 7$ TeV (left) and $\sqrt s = 13$ TeV (right).} 
\label{fig:second}
\end{figure}

Next, in Fig.~\ref{fig:second}, we allow for $Y_A$ (rapidity of the forward jet) and $Y_B$ (rapidity of the backward jet) to take values such that $(Y_A^{\text{min}} = 3) < Y_A < (Y_A^{\text{max}} = 4.7)$ and $(Y_B^{\text{min}} = -4.7) < Y_B < (Y_B^{\text{max}} = -3)$. 
Furthermore, the rapidity of the central jet can take values in five distinct rapidity bins of unit width, that is, $y_i-0.5 < y_J<y_i+0.5$, with $y_i= \{-1, -0.5, 0, 0.5, 1\}$.
Obviously, the ratios now are functions of $y_i$, namely,
\begin{eqnarray}
 R_{PQ}^{MN}(y_i) \, =\frac{ C_{MN}^{\text{integ}}(y_i)}{C_{PQ}^{\text{integ}}(y_i)}\, .
\end{eqnarray}
The $y_i$-dependence of the red, green and blue bin ratio $R^{12}_{33}$ is very weak. From an experimental point of view, this fact is very important as it allows one to select events with the central jet  in a wider range in rapidity, 
leading to better statistics. 
Finally, the similarity between the $\sqrt s = 7$ TeV and $\sqrt s = 13$ TeV plots is more pronounced than in  Fig.~\ref{fig:first} and the NLLA corrections seem to be even softer. 

\section{Rapidity Veto}

\begin{figure}[H]
\hspace{-.9cm}
\begin{subfigure}{.1\textwidth}
\centering
   \includegraphics[scale=0.4]{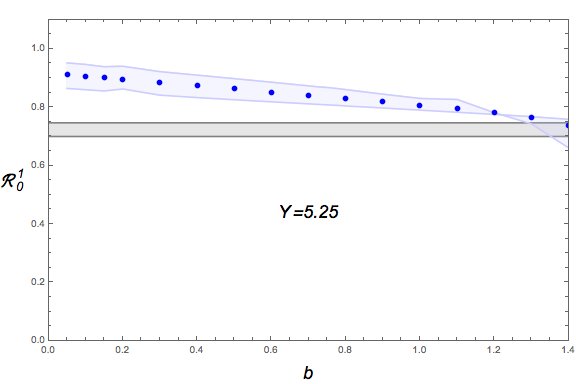}
\end{subfigure}
\begin{subfigure}{1.45\textwidth}
\centering
   \includegraphics[scale=0.4]{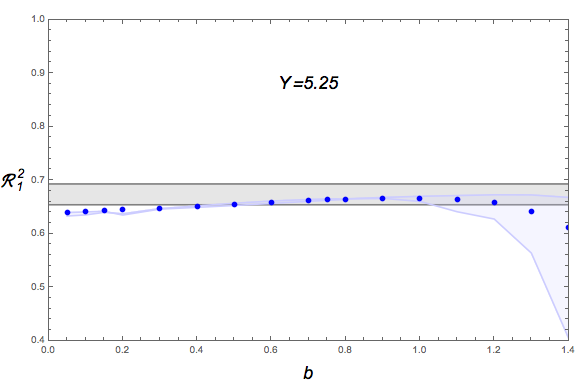}
\end{subfigure}
\caption{\small Veto ($b$) dependence of the azimuthal angle correlation ratios (blue) compared with the experimental value (grey) for $\mathcal{R}^{1}_{0}$ (left) and $\mathcal{R}^{2}_{1}$ (right). For rapidity difference Y=5.25.} 
\label{fig:veto}
\end{figure}

In this section, we report on our work to incorporate a kinematical constraint
in the study of the azimuthal angle correlations in Mueller-Navelet jets. In particular, we choose to impose  
a rapidity veto, that is, we introduce a constraint in the rapidity of the emitted gluons along the gluonic ladder.
The idea, originally proposed by Lipatov in~\cite{Lipatov:1998xx}, is that a significant reduction in the resultant Green function occurs  by demanding that one only considers contributions to the scattering amplitude in which emitted gluons have a minimum rapidity gap, $b$, relative to the preceding emitted gluon.
Further studies on the implications  after including a rapidity veto at NLLA were performed in \cite{Schmidt:1999mz},  focusing mainly on the level of the gluon Green function. Going beyond that level, 
in Fig.~\ref{fig:veto} we show  first results  for two azimuthal ratios (in dijet production) after introducing a rapidity veto $b$ and for various values of $b$. A complete discussion regarding the impact of a rapidity veto in the study of the azimuthal ratios in Mueller-Navelet jets, will be presented elsewhere~\cite{veto_future}.

\section{Summary \& Outlook}

We have shown  results from a first beyond the leading logarithmic accuracy work on generalised azimuthal angle observables in inclusive three-jet production at the LHC within the BFKL formalism. 
For a proper study of all the important corrections beyond the LLA, apart from the NLLA gluon Green functions that we considered here, the NLO jet vertices need to be included since it is expected that they will affect the highest rapidity region where momentum conservation effects are stronger. The most noticeable conclusion from our work so far is that the corrections that come into play after considering NLLA gluon Green functions are generally weak, showing that the generalised observables exhibit perturbative stability. Moreover, we should note that the plots do not change considerably after raising the  colliding energy from 7 TeV to 13 TeV.

In the future, we plan to compare the results presented here against theoretical estimates from fixed order calculations, the full BFKL Monte Carlo \cod {BFKLex}
\cite{BFKLex}~
as well as from general-purpose Monte Carlos tools.

Finally, we have presented an ongoing study where we examine how the imposition of a rapidity veto affects azimuthal correlations of Mueller-Navelet jets. 
The main goal is to establish which values of this parameter are optimal to describe the phenomenology that the LHC exhibits, at current pre-asymptotic center of mass energies.

\begin{flushleft}
{\bf \large Acknowledgements}
\end{flushleft}
This work was supported by the Spanish Research Agency (Agencia Estatal de Investigación) through the grant IFT Centro de Excelencia Severo Ochoa SEV-2016-0597.
GC and ASV acknowledge support from the Spanish Government grants FPA2015-65480-P, FPA2016-78022-P.  
DGG is supported with a fellowship of the international programme ``La Caixa-Severo Ochoa''.
FGC acknowledges support from the Italian Foundation "Angelo della Riccia".

\end{document}